# HEP High Power Targetry Roadmap

**Workshop report**

K. Ammigan[1], C. Barbier[2], S. Bidhar[1], A. Casella[3], M. Calviani[4], C. Densham[5], P. Hurh[1], D. Kim[6], D. Liu[7], K. Lynch[1], S. Makimura[8], F. Pellemoine[1], D. Senor[3], V. Shiltsev[1], D. Stratakis[1], J. Terry[9], K. Yonehara[1]

1 Fermi National Accelerator Laboratory, Batavia IL 60510-5011 USA

2 Oak Ridge National Laboratory, Oak Ridge, TN, USA

3 Pacific Northwest National Laboratory, Richland, WA, USA

4 European Laboratory for Particle Physics (CERN), Geneva, Switzerland

5 Rutherford Appleton Laboratory, Didcot, Oxfordshire, UK

6 Brookhaven National Lab., Upton, NY 11973, USA

7 University of Bristol, Bristol BS8 1TL, UK

8 High Energy Accelerator Research Organization, Tsukuba, Japan

9 Illinois Institute of Technology, Chicago, IL, USA

Table of Contents

# 1 Workshop context and charge

A High-Power Targetry (HPT) roadmap workshop, sponsored by the Department of Energy's Office of High Energy Physics (HEP OHEP), was held April 11-12, 2023, at Fermilab.

The workshop was charged to gather input and information for the purpose of aiding OHEP to plot the course of its Particle Sources and Targets Thrust within the General Accelerator Research and Development (GARD) program.

Information from the workshop would be used to prepare for P5, the Particle Physics Project Prioritization Panel that makes recommendations on the next 10 years of the US particle physics program and would also be useful to HEPAP (High Energy Physics Advisory Panel), that provides long-range (20-year timeframe) advice to DOE and NSF.

This roadmap is envisioned to be helpful to the DOE-OHEP office when planning and prioritizing future R&D activities as well as leveraging synergies across the Office of Science. The roadmap will be extremely beneficial to the broader (external to DOE HEP) HPT community by communicating OHEP's high-level strategy and objectives for HPT R&D and highlighting possible opportunities for collaboration. Results of this workshop build on the extensive community involvement in the Snowmass study.

Development of a robust and reliable high-power target system is of critical importance for many future science experiments and user facilities. The targets must endure high-power CW or pulsed beam, leading to high thermal stresses, pressures, and/or shocks. The cyclic loading environment resulting from the pulsed beam progressively damages the target material's microstructure, often leading to fatigue failure. The increased beam power creates significant challenges such as corrosion and radiation damage that can harm the target materials and degrade their mechanical and thermal properties during irradiation. The integrated damage (thermal stress, thermal shock, fatigue, radiation damage) can eventually lead to the failure of the material and drastically reduce the lifetimes of targets and other beam intercepting devices.

Designing a reliable target is already a challenge for MW-class facilities today and has led several major accelerator facilities to operate at lower-than-design power due to target concerns. With present plans to increase beam power for next-generation accelerator facilities in the next decade, timely R&D in support of robust high-power targets is critical to secure the full physics benefits of ambitious accelerator power upgrades.

The next generation of high-power targets and beam-intercepting devices (beam dumps, absorbers, collimators…) will have more complex geometries, novel materials, and new concepts that allow for use of improved high-heat-flux cooling methods. Advanced numerical simulations need to be developed to support design of reliable high-power beam targets. In parallel, development of radiation-hardened beam instrumentation is needed. Irradiation methods for high-power targets must be further developed, and new irradiation facilities are needed since only a few facilities worldwide offer beams suitable for target testing.

A comprehensive R&D program must be implemented to address the many complex challenges faced by multi-MW beam intercepting devices.

The first day of the workshop included presentations and open discussion on the work already performed during the Snowmass process and was divided into the following R&D topics:

- Context of future HEP facilities
- R&D topic #1: Novel materials and Novel concepts
- R&D topic #2: Modeling needs
- R&D topic #3: Target health monitoring – Radiation-hardened beam instrumentation
- R&D topic #4: Irradiation stations and Post Irradiation Examination (PIE)

The second day of the workshop was dedicated to finalizing the timeline and outlining the roadmap.

The workshop was attended by 19 participants from 9 institutions, the majority of whom were affiliated with HEP facilities. Additionally, several expert representatives from facilities external to HEP, namely Isotope Production and Spallation Sources, also participated to help ensure that the roadmap would be relevant to the broader HPT community.

The agenda and participant list can be found here: https://indico.fnal.gov/event/59123/

# 2 Context of future HEP facilities

During the workshop, those major projects requiring high-power targets that are the most relevant to the HEP HPT roadmap were selected from Figure 2 of the DOE Accelerator and Beam Physics Roadmap (General Accelerator R&D Program, September 6-8, 2022).

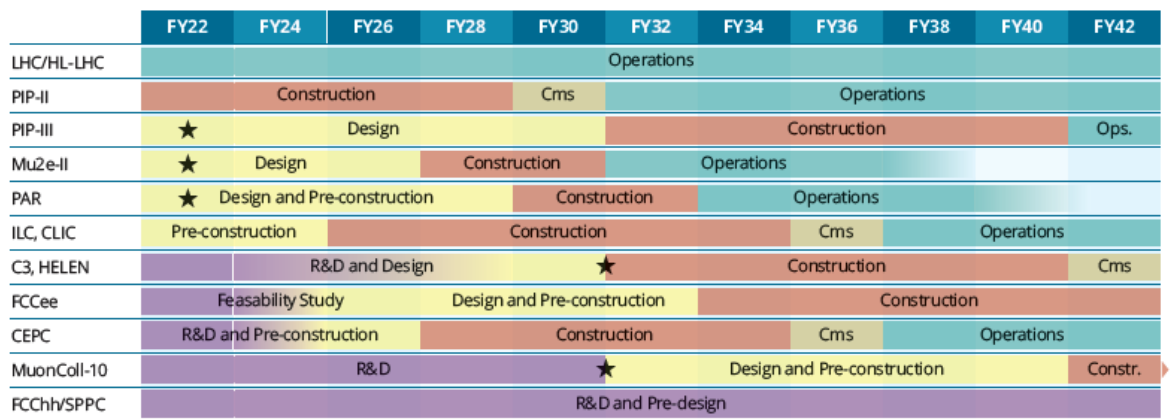

Figure 2: Approximate timeline of HEP accelerators according to the Snowmass'21 and EPPSU reports. *As described in the text, the star symbols indicate major decision points*

Figure 1- Approximate timeline of HEP accelerators according to the Snowmass'21 and EPPSU reports. [1]

As discussed in the workshop, the selected facilities will require reliable production targets and beam-intercepting devices able to withstand multi-MW beams.

Future HEP facilities requiring extensive HPT R&D are listed below in priority order. The list is based on the 2023 P5 report:

## 2.1 Long Baseline Neutrino Facility – LBNF [2]

The Long Baseline Neutrino Facility includes two facilities: the Near Site at Fermi National Accelerator Laboratory (FNAL) and the Far Site at Sanford Underground Research Facility (SURF). Together, these facilities furnish the beam, underground facilities, and infrastructure required to support the Deep Underground Neutrino Experiment (DUNE) [3].

A planned accelerator upgrade (Main Injector Ramp and Target; ACE-MIRT) will provide up to 2.1 MW of proton beam power [4] by the year 2034. The current target design for the early stage of LBNF (2031) can withstand at least 1.2 MW of 120-GeV primary beam, corresponding to 25 kW of power deposited in the graphite target. Simulation and modeling of this target will be used to explore limitations and design margins beyond 1.2 MW, up to 1.5 MW of primary beam power. A second-generation target, based on the initial target design but capable of sustaining up to 1.8 MW of primary beam, is required for the next stage. It is likely that this second-generation target could be realized with minor augmentation of the previous target design. Significant R&D will be required to develop a new target capable of reaching 2 MW or more (or up to 50 kW of beam power deposited in a graphite target) by 2034.

## 2.2 Large circular e+e- collider - FCC-ee [5,6]

FCC-ee is proposed as a phase of CERN's Future Circular Collider. An acceptable FCC-ee target is expected to be made of a high-Z material and should be able to withstand a few kW of beam power [5]. Cooling will be a major challenge for this kind of target due to the significant amount of deposited power. Extensive R&D will be needed to develop a reliable target.

## 2.3 Muon to electron - Mu2e-II [7]

Mu2e-II, using the PIP-II linac, is a midterm evolution of the Mu2e experiment at Fermilab. The current Mu2e target is made of tungsten, radiatively cooled, and can barely withstand 700 W of power deposited by an 8-GeV beam. Mu2e-II will require a 100-kW beam at 800 MeV, corresponding to 20 kW of power deposited in a tungsten target. This power level is extremely challenging for targets made of high-Z materials, and a completely new design may be required. The R&D needs to be carried out as soon as possible to enable appropriate decisions on the target concept to be made, and to have enough time for full design and development of a reliable target.

## 2.4 10-TeV Muon Collider [8]:

The Muon Collider (MuC) promises to be able to extend the lepton-collider energy reach to much higher energies. To produce an acceptable muon current, a 1- to 4-MW proton beam at 5 to 20 GeV is required, with an optimum at 2 MW. Development of a target system capable of reliably withstanding such an intense beam is a substantial challenge. The initial plan is to consider use of solid graphite instead of mercury as the target material. Other options, such as other liquid metals or fluidized targets, will be considered as described in Section 3.

## 2.5 Advanced Muon Facility - AMF [9]

The Advanced Muon Facility (AMF) is a proposal for a next-generation muon facility at Fermilab that would exploit the full potential of the PIP-II accelerator complex to deliver the world's most intense µ+ and µ− beams. This project includes two phased implementations to realize the full potential of AMF: a conceptual design for a primary proton beam of ~100 kW at 800 MeV using PIP-II; followed by a second stage with an upgrade to reach a primary proton beam power of 1 MW. Several target concepts have been developed to withstand 100 kW of primary proton beam power, but significant R&D is needed to design a MW-class target.

## 2.6 PIP-II Accumulator Ring – PAR [10]

The PIP-II Accumulator Ring (PAR) has been proposed to enable the PIP-II linac to better perform two roles; to improve Booster performance and to provide a platform for new HEP experiments.

## 2.7 General R&D to support HEP priority projects.

An extensive R&D program for each of these future facilities is necessary to develop targets able to withstand both high instantaneous current and high average beam power. The muon collider, Mu2e-II, and the first stage of AMF have significant synergies as they face many of the same challenges in target design.

Figure 2 represents the timeline of the future HEP facilities requiring extensive HPT R&D, covering design, prototyping, fabrication, and validation phases of the targets, plus the necessary materials R&D.

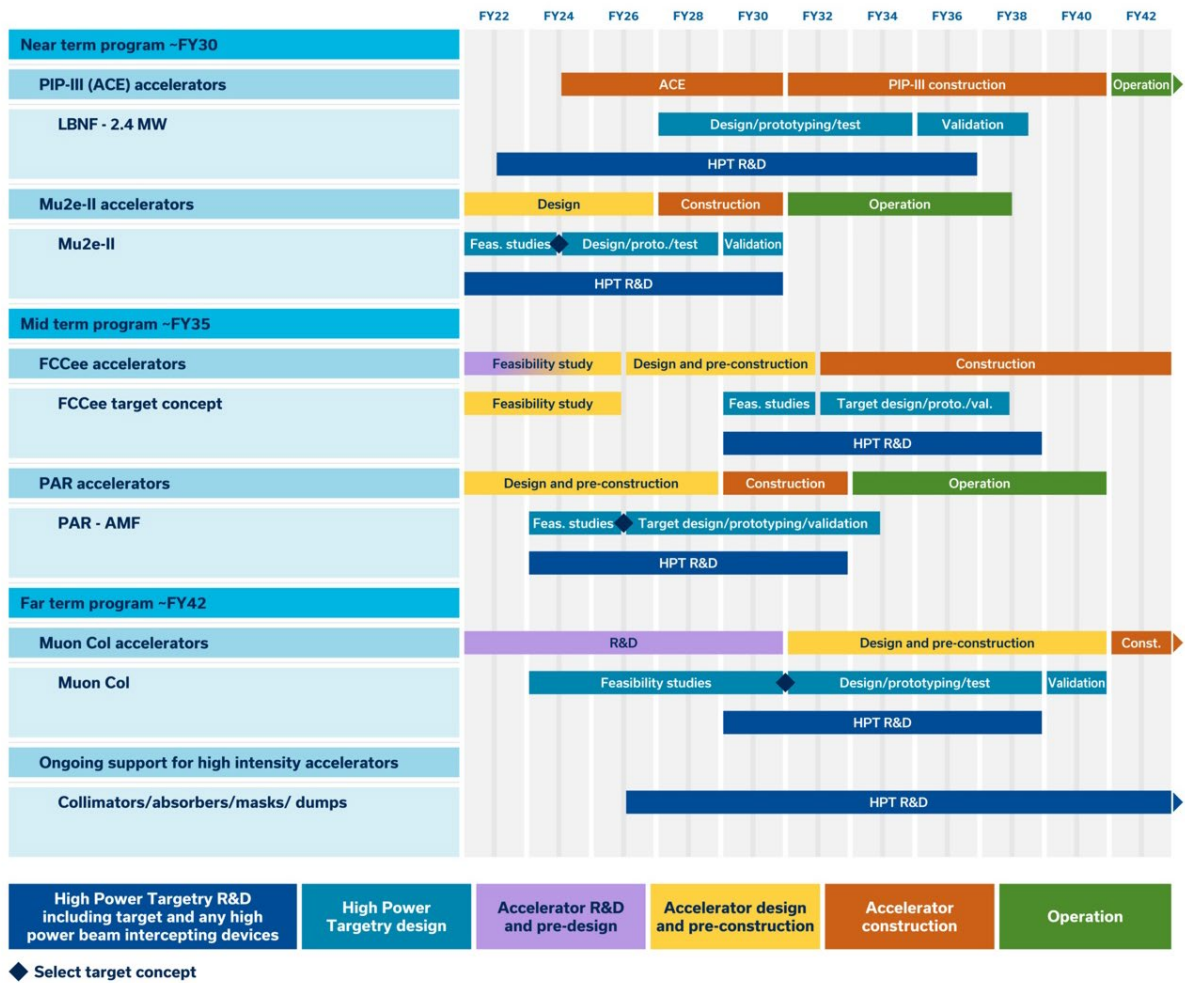

Figure 2 – High-Power Targetry Roadmap for the next two decades, reflecting HEP priorities. HPT R&D support is necessary for target design, prototyping, and testing.

This workshop focused on materials R&D. R&D relevant to the design of a specific target will depend on specific facility and application requirements. Since individual projects do not currently support long-term R&D or fundamental research studies, this roadmap includes an overlap between the target design/technology studies and the early research on materials that will be needed by future projects. Future projects will be able to leverage this work.

The general HPT R&D roadmap presented here needs to be applied to each future project described above.

# 3 New target concepts and technology in the framework of HEP

The next generation of high-power targetry for future HEP accelerators, including beam-intercepting devices such as production targets, beam dumps, absorbers, collimators, and the like, will use more complex geometries, novel materials, and concepts that facilitate high-heat-flux cooling methods. This section presents some of the different concepts and technologies that are used in other facilities and that could be applied in designing future-generation HEP accelerator facilities.

## 3.1 Timeline

Figure 3 represents the timeline to develop potential new concepts and technologies considered for multi-MW targets.

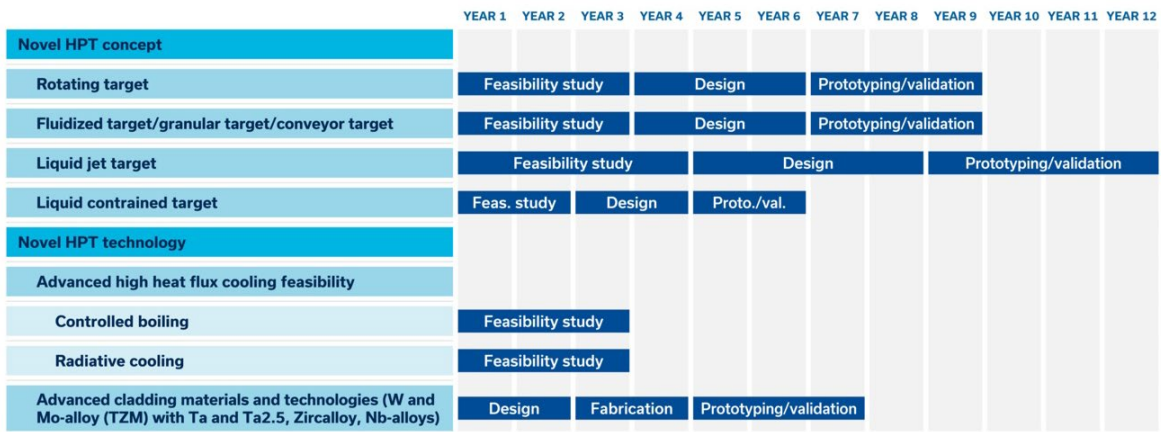

Figure 3 – Timeline of new concept development.

## 3.2 New concepts

Several institutions continue to devote significant effort to the development of current and new targetry concepts and technologies in the context of HEP [11]. These concepts need to be evaluated, and the most relevant explored and further developed over the next several years to address the challenges facing high-power beam-intercepting devices. Significant coordinated R&D efforts in these areas, and synergy between the different projects, will be necessary to enable the safe and reliable operation of future multi-MW accelerator target facilities.

Several of the concepts described below are under consideration, and a feasibility study will inform the selection of the optimal target design. General requirements for production targets are:

- Optimized physics performance,
- Optimized operational temperature,

- Optimized lifetime.

Muon production requires that the target be positioned inside a high-field solenoid. Some of the concepts described here won't be realizable in such a high-magnetic-field environment but may be useable in other applications or other types of facilities.

### 3.2.1 Rotating target

An option that is commonly used to reduce the power density of beam hitting a target is to rotate the target. In rotating target applications, the incoming particle beam is held stationary while the target is rotated. Rotating targets have been developed and are being optimized to spread the heat deposition footprint and thus minimize radiation damage to the target material. Some of the associated challenges include development of a reliable mechanical system to drive the rotation, including design of robust, radiation-hard bearings and lubricants, to help enable operation in high-radiation and high-temperature environments.

### 3.2.2 Liquid target:

**Liquid-contained target**: The SNS and J-PARC use circulating liquid mercury targets; PSI performs research using a liquid lead-bismuth eutectic as the target material. These targets exhibit excellent power-handling capability. Target vessels containing circulating liquids are used when fixed solid targets cannot handle the power density of the primary beam. The advantage of circulating liquid targets is that heat deposited in the fluid is rapidly swept away, allowing for a much higher thermal load since fatigue failure in the core of the liquid target is not an issue. That said, the beam windows integrated into liquid target containment vessels are exposed to high heat deposition, mechanical fatigue, shock, radiation damage, cavitation-induced pitting, and other damage due to the beam. Significant R&D is needed so that future designs will be able to overcome these complex and challenging issues. Safety concerns related to the potentially-highly-activated liquid target material must be addressed in the design of the target and handling equipment.

**Liquid-jet target**: Liquid-jet targets have the advantages of the liquid-contained target but are designed to operate without beam windows. A liquid-jet target concept was demonstrated by Brookhaven National Laboratory and Princeton University [12]. The MERIT (nTOF-11) experiment at CERN demonstrated that the technology is adequate for beam power up to 4 MW inside a 15-T magnetic field [13]. Safety concerns related to the potentially-highly-activated liquid target material must be addressed in the design of the target and handling equipment.

### 3.2.3 Flowing target:

**Fixed granular target design**: This type of target consists of a matrix of granular tungsten cooled by flowing gaseous helium. Such a design is compact, but its high radiation damage would likely require frequent replacement of the target.

**Conveyor concept:** This concept [14] is based on spherical tungsten or carbon balls that are inserted into a pipe, transported to the beam interaction area, and then transported away for cooling and replacement (as necessary). This design would occupy a relatively small space (consistent with the Heat and Radiation Shield (HRS). Helium gas could be used for both cooling and moving elements inside the conveyor's pipe. Radiation damage can also be distributed among a large number of replaceable elements. The design of such a system is technically complex and will require prototyping and extensive testing.

### 3.2.4 Existing concepts

Table 1 lists existing concepts as described in the section above with their limitations and current status.

| Concepts | Institutions | Comments | Limitations | status | Ref |
|---|---|---|---|---|---|
| Rotating C target | Facility for Rare Isotope Beams | 5000 rpm, 400 kW heavy ion beam (100 kW of deposited power). | Radiation resistance of rotation system component (bearing, motor) | Under Operation at low power (10 kW), demonstrated for full power with electron beam | [11] |
| Rotating W target | European Spallation Source | 5 MW pulsed proton beam | Radiation resistance of rotation system component (bearing, motor), survivability of the cladding component | Under fabrication | [11] |
| Liquid Hg target | Spallation Neutron Source | Designed to operate up to 2MW of proton beam power | Cavitation, weld aging | Operates up to 1.5 MW | [11] |
| Liquid Jet target | MERIT experiment | Technology demonstrated at CERN for operation up to 4 MW | Only at a concept level | Technology tested | [11], [13] |
| Powder fluidized target | Rutherford Appleton Laboratory | Concept with W powder tested under pulsed beam test at CERN-HRMT facility | Only at a concept level | Concept level, to be validated | [11] |
| Conveyor target | Fermilab - Mu2e-II | Designed for 100 kW beam (30 kW of deposited power) | Only at a concept level | Concept level, to be validated | [14] |

Table 1 – List of High-Power Target concepts, either operating or being designed.

## 3.3 New Technologies

### 3.3.1 Advanced cladding materials and technologies:

Refractory metals are widely employed in laboratories worldwide for secondary particle production, such as for neutron production [15] or for proposed beam dump experiments [16], owing to their reduced nuclear interaction length and relatively low neutron inelastic cross-section. However, direct cooling with water is not possible due to the high hydrogen embrittlement. Usually, tungsten (W) is cladded via Hot Isostatic Pressing (HIP) with pure Tantalum (Ta), with good results [17]. Advanced techniques to clad pure W and other refractory metals with Ta and Ta2.5W are being developed [18] [19] [20]. For high-power facilities, decay heat on Ta and Ta-alloys may pose safety and other concerns, and therefore other cladding materials such as Zircalloy or other Nb-alloys are being studied. Efforts to study and improve cladding technologies (e.g. Hot Isostatic Pressing and diffusion bonding) are ongoing to mitigate cladding breach, corrosion, and erosion issues. R&D in advanced cladding materials and technologies will be beneficial to enable and optimize future experiments [15] [16]. It will be fundamental in designing robust beam-intercepting devices for planned upgrades and intensity and energy frontier facilities of the future.

### 3.3.2 High-heat-flux cooling:

Exploration and development of alternatives to forced-convection cooling techniques is urgently needed, as these alternatives will be critical in enabling future multi-MW target facilities. Unconventional heat transfer technologies, such as controlled boiling or flowing (liquid or granular) targets, where heat-flux cooling capacity can be significantly increased, need to be explored and developed. High-heat-flux cooling techniques have been investigated in the past and continue to be researched within the fusion/fission communities, as well as in some accelerator target facilities. Optimized radiative cooling is another technique that could be used in some applications, such as in the Mu2e tungsten target.

## 4 Target materials

The development, characterization, and qualification of materials suitable for high-power targetry facilities is required. It is imperative to proactively engage in materials research and development ahead of the design phase, given the extensive time required for the R&D work. The earlier we can determine the selection of a particular target concept for a specific facility, the sooner we can begin the essential materials R&D activities.

The use of conventional materials for key accelerator components, such as beam windows and secondary-particle-production targets, already limits the scope of today's experiments.

This section addresses development of the kinds of novel target materials that can enable new target concepts to be realized. Novel target materials will help pave the way for future multi-MW beam operation, maximizing the physics potential of future HEP experiments.

### 4.1 Timeline

Figure 4 shows the possible target materials under consideration by the various selected facilities.

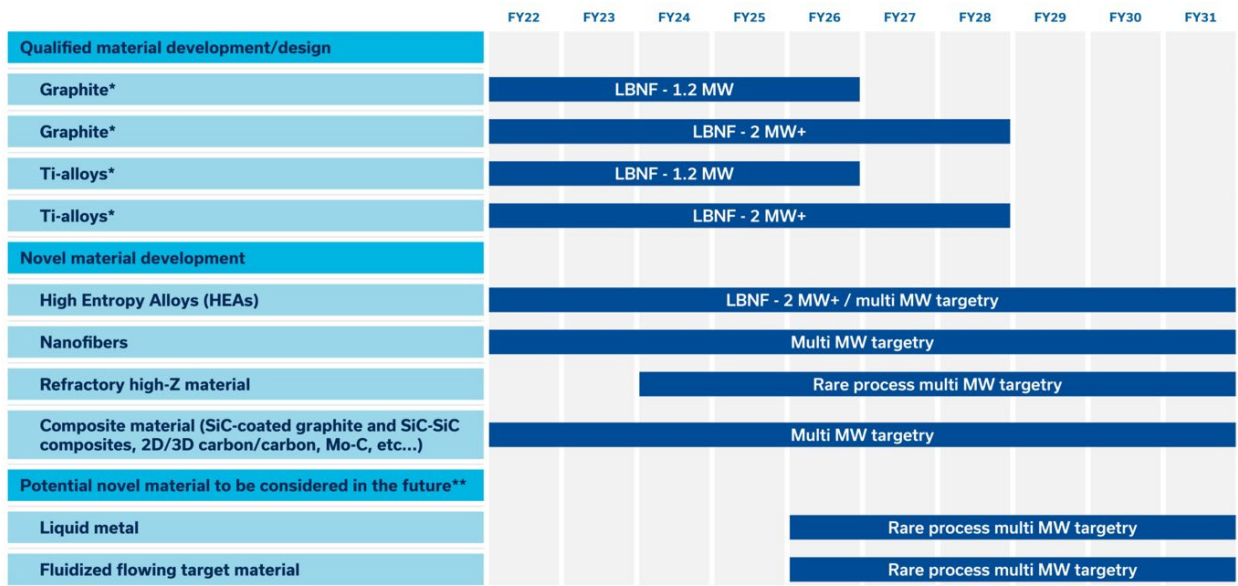

Figure 4 - Anticipated materials related to the selected facilities.

## 4.2 Novel Materials

The goal of new alternative-material development is to enhance the performance, reliability, and operational lifespan of next-generation multi-megawatt accelerator target facilities. These efforts will specifically address the leading cross-cutting challenges posed by beam-induced radiation damage and thermal shock.

Controlling the severity of thermal shock in materials involves targeting the material's microstructure and its inherent capacity to withstand thermal expansion and dampen stress waves. This can be achieved by developing a material with lower density, elastic modulus, coefficient of thermal expansion, along with higher specific heat capacity and yield strength [21].

The impacts of radiation damage can be alleviated by producing extremely fine grains and high-density grain boundaries in the material's microstructure. Another strategy involves the development of materials that are intrinsically resistant to radiation damage. The overarching objective is to minimize undesirable outcomes such as embrittlement, hardening, swelling, reduction of thermal conductivity, and phenomena dependent on diffusion such as the segregation of impurities and phase transformation – all of which pose threats to the integrity of targets.

Some of the most promising novel material candidates that need to be explored and developed for specific beam-intercepting devices are described below. R&D is necessary to determine whether they are capable of withstanding increased beam power and intensity. Most of the R&D efforts described below will continue to be coordinated within the RaDIATE Collaboration [22].

### 4.2.1 High-Entropy Alloys (HEAs)

High-Entropy Alloys (HEAs) are fundamentally different from conventional alloys, and typically consist of several principal elements in near equimolar quantities. The resulting alloys are materials with structures and properties that are not dictated primarily by a single element, but rather behave as an average of each primary constituent. Research into HEAs over the last decade has shown that this class of materials exhibits a broad range of promising properties for both structural and functional materials, including high yield strengths, fracture toughness, and oxidation resistance, even at elevated temperatures. The most relevant and intriguing property of HEAs for accelerator target applications is their microstructural response to radiation damage (from [23] to [30]). HEAs offer a unique opportunity to explore a broader range of novel radiation-damage-resistant alloy systems with functional properties specific to accelerator beam-intercepting devices, such as beam windows.

### 4.2.2 Electrospun nanofiber materials

Nanofiber materials are promising for future multi-MW targets as they are intrinsically tolerant to both thermal shock and radiation damage. Since the continuum is physically discretized at the microscale, issues such as thermal shock, thermal stress cycles, and local heat accumulation can be mitigated. Owing to the nanopolycrystalline grains of individual nanofibers, it is hypothesized that they would also offer better resistance to radiation damage. The large number of grain boundaries and free surfaces would act as sinks to irradiation-induced defects. Initial work on ceramic nanofibers has revealed evidence of radiation damage resistance upon low-energy heavy ion irradiation [31]. However, more systematic studies are needed to qualify potential metallic nanofiber materials for targetry applications.

### 4.2.3 Refractory high-Z materials

Tungsten (W) is a principal candidate target material because of its high density and extremely high melting point, and because it can provide 10 times higher brightness of muons/neutrons than that of current target materials [32]. However, W is already brittle at room temperature and exhibits significant embrittlement due to the recrystallization that occurs when W is heated at or above the recrystallization temperature (almost one third of the melting point). In addition, significant embrittlement is observed from proton irradiation damage [33]. TFGR (Toughened, Fine Grained, Recrystallized) tungsten alloy, originally developed at Tohoku University and now continued by KEK, has been shown to overcome the embrittlement issue due to grain-boundary-reinforced nanostructures [34] [35]. Further R&D work is ongoing at KEK.

### 4.2.4 Composite materials

Graphite shows extremely good performance when used in proton beam target applications due to its thermal and mechanical properties and chemical stability. However, graphite oxidizes easily at high temperatures, and oxidation contaminants can complicate recovery procedures and downtime access in accelerator target facilities.

Work is currently ongoing to study Silicon Carbide (SiC) coated graphite to improve oxidation resistance. Another promising target material candidate is Nano-powder Infiltration and Transient Eutectoid (NITE) SiC/SiC composite, as it is significantly denser and can provide higher-efficiency secondary-particle transport. The composite also exhibits high oxidation resistance and pseudo-ductile behavior, enabling it to withstand higher stresses (from [36] to [39]).

Advanced graphitic materials such as 2D carbon/carbon and 3D carbon/carbon and different novel grades also need to be explored as these materials have high temperature resistance, low density, and low thermal expansion coefficients. More importantly, they also show excellent performance under radiation due to the crystal lattice configuration, and they have good annealing capabilities. Some of these materials have already been employed in accelerator facilities. They were also tested at CERN's HiRadMat facility to demonstrate their capability to withstand extreme operational conditions [40].

## 5 HPT Material R&D program

Radiation damage, thermal shock, and thermal fatigue are identified as the most cross-cutting challenges of high-power target facilities and need to be addressed in an extensive material R&D program.

This section addresses the general scope to perform a complete material R&D program in the framework of High-Power Targetry development, including the three challenges mentioned above.

### 5.1 Timeline

The general Roadmap and timeline for developing target materials for specific applications and facilities are described below, with an overview shown in Figure 5.

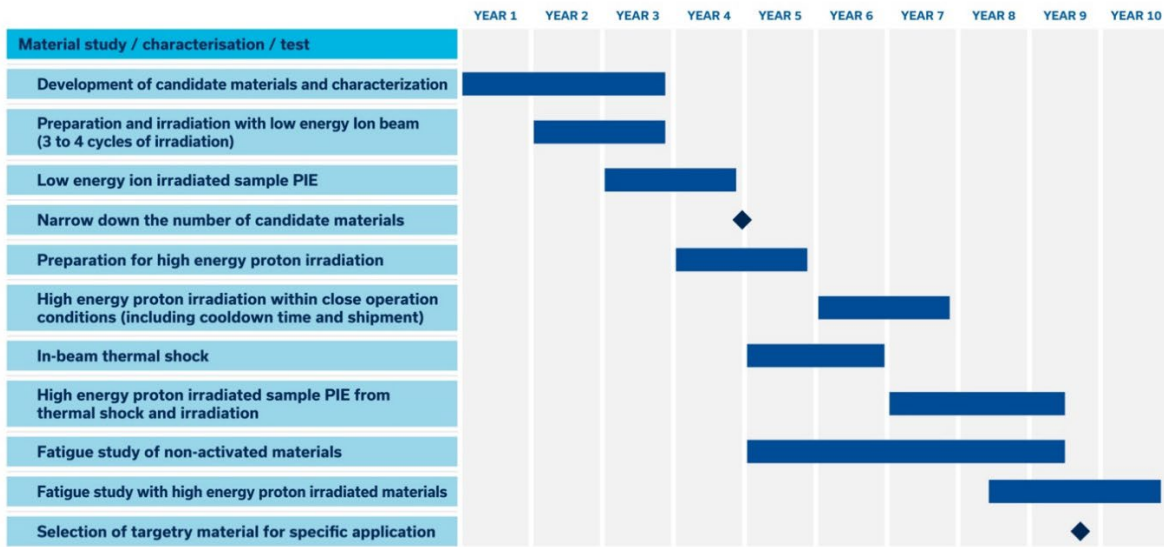

Figure 5 – Timeline for an extensive HPT material R&D cycle.

## 5.2 Development and manufacturing of candidate materials and characterization

This activity will focus on the preparation of different grades of materials with different heat treatments, components, and fabrication methods that will impact their physical properties and resistance to radiation damage. This phase will include simulations to define material components and extensive characterization to measure their physical properties.

## 5.3 Low-energy ion irradiation

Exposure of material to particles leads to a displacement of a significant number of atoms from their preferential lattice (quantified as displacements per atom, DPA, a reference used to express radiation damage level in material), transmutation and gas production. Irradiation-induced defects such as dislocation loops, point defect clusters, fine-scale precipitates and voids that accumulate at the microstructural level ultimately disrupt the lattice structure of the material and affect the mechanical, electrical and other physical properties of irradiated materials. Production of hydrogen and helium can lead to drastic changes in physical properties and must be carefully studied.

Directly studying material properties of targets exposed to high-energy proton accelerators is expensive and time-consuming due to the high levels of activation and the low radiation damage rate. The highly-activated samples require characterization in hot cells equipped with advanced test machines. These facilities are typically expensive to use, and the low availability of beam time is restrictive.

Use of low-energy (LE) ion irradiation could possibly address limitations of high energy proton beam irradiation, providing an effective and fast way to produce radiation damage in materials without activating the specimens. A characteristic feature of radiation damage by high-intensity proton beams is the formation of helium and hydrogen by nuclear transmutation in the material. Irradiation with low-energy heavy ions cannot produce transmutation gas. Use of ion beam facilities with combined beamlines: dual or triple beams, will allow for damage creation by atom displacement (dpa) from a heavy-ion beam coupled with implantation of helium and/or hydrogen by the other beam(s). It can then be possible to investigate the synergistic effects of damage and to also reproduce transmutation-gas production in the material. We will use this method to effectively explore the candidate materials for future high-intensity target facilities. Three to four irradiations can be feasible during this two-year phase.

## 5.4 Low-energy-ion irradiated sample Post Irradiation Examination (PIE)

Low-energy ion irradiation is a more cost- and time-efficient method to screen candidate materials, even with its limitations, than the more cost- and time-demanding high-energy proton irradiation. One drawback to the use of low-energy ions is the shallow and non-homogeneous implantation depth (between 1 to a few micrometers depending on ion energy and material). A limitation is that microscale characterization techniques don't necessarily correspond to the physical properties of the material on a macroscopic scale [bulk]. However, these techniques can effectively probe the shallow damage region to measure relative changes in thermal and mechanical properties before and after irradiation. Despite this limitation, low-energy heavy-ion irradiation is very attractive and microscale techniques need to be developed to extend the assessment of radiation damage effect in materials.

## 5.5 Down-select to the best candidate materials:

The radiation damage obtained by low-energy irradiation is restricted to a depth of few microns from the surface of the material and can only be measured with microscale testing. A change in hardness can be used as a clear indicator of damage to the material and can be obtained from nanoindentation techniques.

Microscale test results on thermal and mechanical properties of irradiated and non-irradiated samples will be used to help select a small number of the best candidate materials. These samples can be used to proceed to prototypic irradiation under conditions closer to those of actual beam operation.

## 5.6 High-energy proton irradiation

Use of high-energy proton beam irradiation will lead to the production of activated materials, so careful planning is required. The irradiation duration depends on facility capabilities (beam intensity, and beam energy) and schedule (we may need to irradiate the specimen during different beam periods due to facility constraints). This period could range from 2 weeks to a few months. The timeline in Figure 5 includes a cooldown time that depends on how activated the specimen is after irradiation.

## 5.7 In-beam thermal shock

During operation, targets need to sustain intense thermal shocks due to the high-intensity beams. Single pulsed-beam thermal-shock tests were performed in the past at the CERN-HiRadMat Facility to understand the failure mechanisms, limits, and flow behavior of the various material specimens (both irradiated and non-irradiated).

Similar tests need to be pursued with new materials and results of these tests will inform selection of the best candidate materials. Irradiated samples can be taken from a used target or from a target sample that was subjected to high-energy proton irradiation as described above.

After each campaign (either irradiation or thermal shock) samples will be sent to the appropriate facilities for Post Irradiation Examination (PIE). PIE techniques are described in more detail in Section 9.

## 5.8 Fatigue study

### 5.8.1 Fatigue study of non-irradiated materials:

The candidate materials will be tested for high-cycle fatigue before irradiation with high-energy protons. A conventional mechanical fatigue test machine requires a lot of time to build the S-N curve and reach the 10s of millions of cycles expected during operation. In practice, it is recommended to conduct tests with at least 4 or 5 stress levels above the assumed fatigue limit and to conduct 6 to 15 tests at each stress level.

### 5.8.2 Fatigue study with irradiated samples:

Similar tests need to be performed with irradiated samples, but shapes and sizes of those specimen are non-standard (ASTM). A special shape will be designed to accommodate the irradiation experiment constraints. A special fatigue-test machine that can be used in a hot cell with activated samples will be built to test those specimens.

## 5.9 Selection of candidate material for a specific application:

A selection of the most promising materials for each application (beam intercepting device and specific facility) will be made by comparing the results from the different studies performed during the last 5 to 10 years. Studies considered will include high-energy irradiation, in-beam thermal shock, and fatigue.

This R&D cycle will only be possible with continued development of other research topics:

- Modeling development,
- Instrumentation development,
- Alternative methods development,
- Post-Irradiation Examination development.

These research topics will be described in the following Sections.

# 6 Modeling development

Simulations play a critical role in design of targets and other accelerator components such as beam dumps, collimators, absorbers, scrapers, and the like. Combined with experimental validation tests, simulations can predict beam-matter interactions, thermo-mechanical responses induced by beam heating, evaluate radiation damage as a function of time, and help ensure safe operation. The next generation of high-power targets will use more complex geometries, novel materials, and possibly entirely new concepts. Current numerical approaches are insufficient to enable convergence to a reliable target design that satisfies the physics requirements. This section discusses necessary improvements in target modeling over the next decade (see Figure 6) to support MW-class targets.

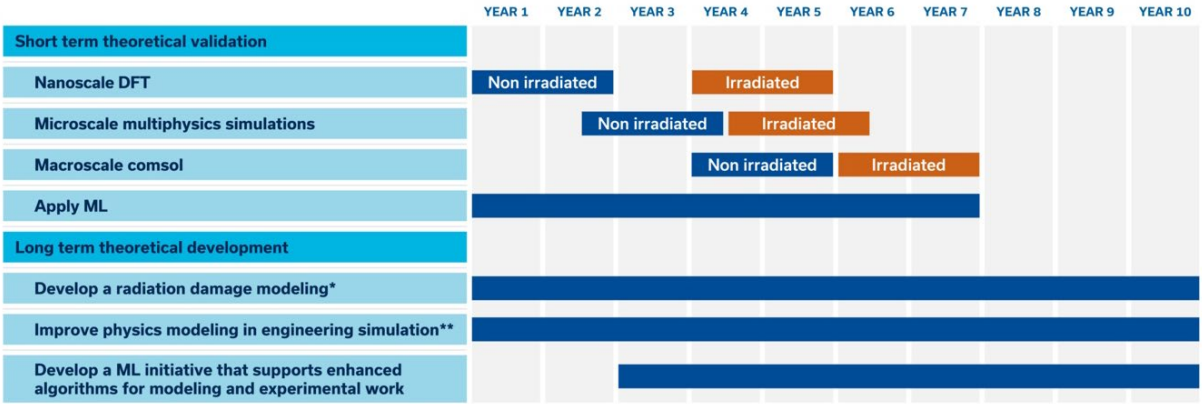

Figure 6 –Timeline for Modeling development to support HPT materials R&D programs.

The physics in a high-power target is complex and requires unique modeling expertise. Beam interaction calculations are required to estimate the target performance and to provide the heat load for thermo-mechanical simulations. Structural analyses to estimate thermal stresses, fatigue, and response to beam pulses (for pulsed-beam applications) are then performed to validate the mechanical integrity of the target. If the target is cooled by a fluid, computational fluid dynamics analyses may also be performed. The above methodology has led to successful and reliable target designs in the past, but limitations are becoming evident in MW-accelerator applications. As the beam power increases, targets face several challenges, including high energy densities, high power densities per pulse, and high average deposited power. A primary difficulty is to provide adequate cooling of the target and its supporting structures. As described in Section 3, one solution is to use a fluid target such as liquid metal or a fluidized material (granular or pellets) to facilitate heat transport. Another solution is to rotate the target to spread the heat load over a larger volume. Both solutions are challenging to model due to their more complex geometries, and to the additional physics introduced (like cavitation in liquid metal targets).

As the beam power increases, radiation damage will limit the target life span. No available numerical engineering tool can model the material degradation (radiation hardening, embrittlement, swelling, change in material properties). Empirical data are often used but can be very limited for some materials and do not allow inclusion of novel materials. To help ensure that operating costs of targets and other beam intercepting devices remain reasonable in future MW facilities, progress in modeling must be made and is described below [41].

When a material is irradiated with energetic particles such as electrons, heavy ions, or protons, the atoms in the material experience displacement from their lattice sites. These displacements cause microstructural defects. Material composition and properties can also be altered by nuclear transmutations. More precisely, displacements can cause swelling, hardening and embrittlement, and can affect structural stability, thermal, electrical, and optical properties. The current standard to estimate radiation damage is to use particle codes with variables such as DPA (Displacement Per Atom). However, these codes do not predict the changes in the material's mechanical properties, and target designers must use empirical data to understand how the material may be impacted by irradiation.

Development and selection of target materials requires a deep understanding of the relation between the change in mechanical properties due radiation damage and what occurs in the material at the microscale during irradiation. Gaining this knowledge is extremely important in the framework of an HPT R&D program.

# 7 Instrumentation development

Development of radiation-hardened beam instrumentation is crucial to facilitate the successful operation of a multi-MW accelerator system. Reliable beam instrumentation is needed for monitoring and control of accelerator operating conditions, and to enable fault tracing of accelerator components including targets [42] in case of problems.

Required instrumentation will be specific per application/facility. This Section addresses instrumentation to be used for monitoring, prototype testing, demonstration, and validation of any HPT system.

## 7.1 Timeline

| | YEAR 1 | YEAR 2 | YEAR 3 | YEAR 4 | YEAR 5 | YEAR 6 | YEAR 7 | YEAR 8 | YEAR 9 | YEAR 10 |
|---|---|---|---|---|---|---|---|---|---|---|
| **In-situ instrumentation** | | | | | | | | | | |
| **Target health monitoring instrumentation development** | | | | | | | | | | |
| Light optics including image visualization detector | Investigation | | R&D/prototyping | | Validation | | | | | |
| LDV, strain sensor | Investigation | | R&D/prototyping | | Validation | | | | | |
| Thermocouple sensor, IR monitor, acoustic sensor | Investigation | | R&D/prototyping | | Validation | | | | | |
| Investigate new technology | ⟶ | ⟶ | ⟶ | ⟶ | ⟶ | ⟶ | ⟶ | | | |
| **Primary beam monitoring instrumentation development** | | | | | | | | | | |
| SEM | Investigation | | R&D/prototyping | | Validation | | | | | |
| OTR | Investigation | | R&D/prototyping | | Validation | | | | | |
| BLM | Investigation | | R&D/prototyping | | Validation | | | | | |
| Luminescence beam profile | Investigation | | R&D/prototyping | | Validation | | | | | |
| ASIC | Investigation | | R&D/prototyping | | Validation | | | | | |
| Investigate new technology | ⟶ | ⟶ | ⟶ | ⟶ | ⟶ | ⟶ | ⟶ | | | |
| **Secondary beam monitoring instrumentation development** | | | | | | | | | | |
| Ion chamber, EMT, CT, LGAD, gas filled RF | Investigation | | R&D/prototyping | | Validation | | | | | |
| ASIC | Investigation | | R&D/prototyping | | Validation | | | | | |
| Investigate new technology | ⟶ | ⟶ | ⟶ | ⟶ | ⟶ | ⟶ | ⟶ | | | |
| Implement instrumentation (Identify physics object, design the system) | ⟶ | ⟶ | ⟶ | | | | | | | |
| **Ex-situ instrumentation** | | | | | | | | | | |
| Non Destructive and fast Test development to monitor health target off beam | ⟶ | ⟶ | ⟶ | ⟶ | ⟶ | ⟶ | ⟶ | ⟶ | ⟶ | ⟶ |

Figure 7 -Timeline for Rad-Hard instrumentation development to support HPT systems.

## 7.2 Beam monitoring devices

A beam monitor is used to characterize the beam by measuring the spatial distribution of charged particles (e.g. beam centroid position, orientation, and profile), the integrated charge of particles passing through the monitor (e.g. a total beam intensity per spill), and the time structure of charged particles (e.g. a differentiated beam intensity or a bunch structure) along the beam line. Optical Transition Radiation (OTR) monitors and Beam Induced Fluorescence (BIF) monitors have been developed and are used at various accelerator facilities. The technology used in these detectors is suitable for a multi-MW beam because it uses a highly sensitive light detector that requires only a small intercept of the beam to measure the beam profile. Determining light yield and acceptable thermal stress on a radiator are key aspects of R&D. Because only a small amount of light transmits with the beam related information, maintaining high light transmission efficiency is crucial for the system performance.

Secondary Electron emission Monitors (SEM) are potential candidates for intense beam facilities. R&D is required to investigate the effect of aging on the Secondary Emission Yield (SEY) and to explore the challenge of creating a thin wire or film to minimize the beam interruption. New SEY materials, like Ni, carbon graphite, and Carbon Nano Tube (CNT), are being considered for multi-MW accelerators. The ionization chamber is a mature technology if beam interception by the detector is not an issue. Due to its simple structure, it can function as a highly radiation-tolerant detector when designed carefully. Radiation-resistant ceramic or plastic is used as an electric insulator, and signal saturation due to space charge is often addressed by selecting the right ionization gas (mixing dopant) and adjusting the dimensions of the electrodes.

The Beam Loss Monitoring (BLM) system is another crucial component of beam instrumentation. It is used for beam diagnostics to tune beam optics parameters by monitoring loss patterns and locations. Additionally, the BLM system plays a key role in machine protection. CERN has developed a radiation hardened Beam Loss Monitor (BLM) system for HL-LHC. The ionization chamber is commonly used for BLM. CERN also developed a silicon BLM with a faster response time than the ionization chamber.

Secondary and tertiary beam profile monitors have been developed for high energy beam applications and are used to characterize the secondary and tertiary particles from the target. The phase space evolution of fragmentation particles in the transport line needs to be understood. The signal is sensitive to the health of the target system and the primary beam condition. SEM and gas-RF monitor [42] are considered as hadron monitors. Electron Multiplier Tubes (EMT) [43] and Current Transformers (CT) are new candidates and were demonstrated at J-PARC as muon monitors. They appear to have a good stability with respect to the beam intensity. Additional R&D is needed for multi-MW beam applications.

## 7.3 In-situ target health sensor

In-situ target health sensors are widely used to diagnose the target condition and to predict the operational lifetime of the target in medium-energy nuclear facilities. However, this type of sensor is not common in high-energy physics applications because it can be an obstacle for the secondary and tertiary particles. We are considering either making a compact in-situ sensor or developing a non-contact sensor for high-energy physics applications.

Fiber-optic strain sensors are used to measure MHz dynamic strains on the target vessel induced by individual proton pulses. Acoustic sensors and Laser Doppler Vibrometers (LDV) are utilized to measure a specific vibration mode in the target system that is generated by beam impact. Additionally, thermocouple sensors are used to measure the temperature distribution in the target system and indicates the beam profile. Frequently, the thermocouple sensor serves as a machine protection detector to protect the target system.

We are also considering a non-contact (ex-situ) target health sensor for high-energy physics targets. Infra-Red (IR) detectors can be used to remotely detect a thermal distribution in the target system.

Luminescence light from Cr doped Alumina is widely used to measure the beam profile and intensity on the target. The emitter covers a large area for a large spot-size beam and measures the 2D profile accurately, even at the beam halo. Mitigating the aging effect is a key R&D for this type of device. Additionally, we are exploring a means to directly measure radiation damage on the material surface by monitoring the refractive index. These non-contact sensors will be tested in the high energy targets. They utilize various types of light optics to extract the light signal from a highly-radioactive environment and transport it to a low radiation area. The lifetime of the light optics is a concern and is described in the next section.

### 7.4 Radiation hardened light optics and light sensor.

The common critical elements in radiation-hardened beam instrumentation are the light guide (e.g. an optical fiber) and optical elements (e.g. a lens and mirror). These light optics are widely used to inject probe light or extract optical measurement light from a beam line and target. Mitigating radiation-induced attenuation (RIA), which indicates a lifetime of an optical material as a function of an integrated radiation dose, poses a challenge. Because the optical measurement light transmits the beam-related information, the refractive index profile of the fiber as a function of the integrated radiation dose becomes crucial. High-purity silica and Fluoride-doped optical materials appear to have higher radiation tolerance than standard silica optics [44].

It is also crucial to develop a radiation-hardened image sensor. While radiation-hardened CCD cameras were once used in many applications, some manufactures have unfortunately reduced their production due to a shrinking market. Several institutions have reported replacing CCD camera with CMOS or CID cameras. The J-PARC neutrino group continues to develop a CID image sensor for the Optical Transition Radiation system. The collider detector group has demonstrated a radiation-hardened CMOS that performs in the interaction region. Generally, CMOS spatial resolution and data acquisition speed are better than the CCD, while the radiation tolerance of the CCD is higher than that of the CMOS or CID. We propose continuous efforts to develop radiation-robust CMOS.

### 7.5 Radiation hardened ASICs

Detectors and readout electronics are unit structures that are exposed to extremely high radiation doses. The CERN and Fermilab collider groups have developed advanced technology for radiation-resistive Application-Specific Integrated Circuits (ASICs) to readout and manipulate signals from a high-granularity fast detector in an extreme environment for the HL-LHC collider detector and future collider detectors. [45]. A beam halo detector has been demonstrated in the LHC and is one example of a radiation-hardened ASIC. These technologies will be adopted for making a new type of fast beam monitor system.

## 8 Alternative method development

Directly studying material properties of targets exposed to high-energy proton beams is an expensive and time-consuming method due to the high levels of activation and the low radiation damage rates. The high activation of samples requires characterization in hot cells equipped with advanced test machines. These facilities are typically expensive to use, and beam time availability is restrictive.

In this Section we will discuss alternative methods for more cost- and time-efficient studies.

## 8.1 Timeline

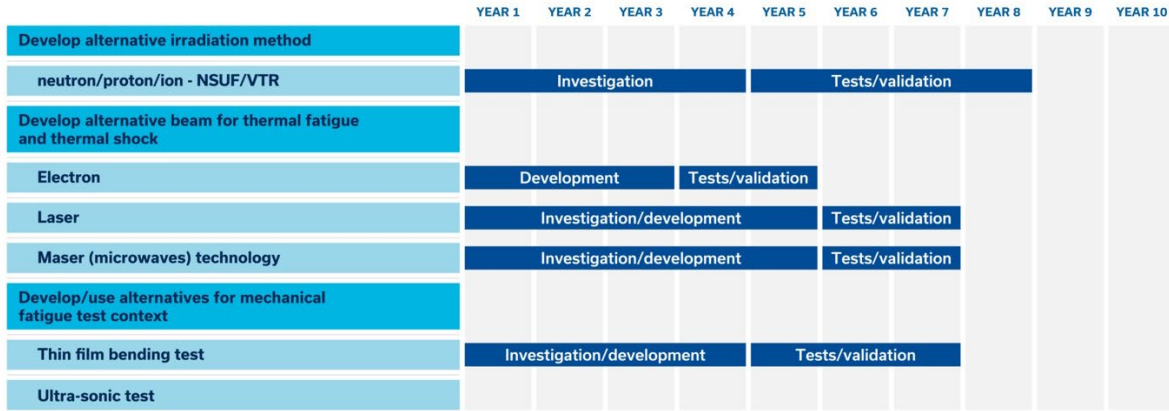

Figure 8 - -Timeline for alternative method development to support HPT material R&D program.

## 8.2 Develop alternative irradiation method.

As mentioned in Section 5, low-energy ion irradiation will address some of the issues related to high-energy proton beam irradiation damage, while providing a rapid and effective way to produce radiation damage in materials without activating the specimens. Low-energy ion irradiation also facilitates exploration of candidate materials for future high-intensity target facilities. Comparing neutron-induced radiation damage to that from high-energy protons could also be investigated, as there is a large amount of published experimental data for several materials of interest to target designers. Handling the highly-activated samples will still require PIE facilities with hot cells that are equipped with limited thermo-mechanical tests.

## 8.3 Develop alternative beam to study thermal fatigue and thermal shock

The method currently used to test a material's response to the localized thermal shock expected in real operation is radiation by an intense single pulsed proton beam at the CERN-HiRadMat facility. Alternatives are currently being explored to use intense proton sources, electron sources, or laser sources to test thermal shock response more effectively and enable high-cycle fatigue studies ($> 1 \times 10^7$ loading cycles). The advantage of using electron or laser beams is that testing using prototypical beam loading parameters leads to no activation of the specimens. Free-electron masers (FEM) were used to study thermal fatigue of an accelerator cavity surface within a penetration depth of several to tens of micrometers [46]. FEM techniques produce mechanical stresses equivalent to those induced by pulsed heating, but the stresses propagate only few micrometers in depth. Results obtained were well correlated with fatigue studies (S-N curves) obtained by other techniques such as Ultrasonic fatigue testing. Use of FEM is also appropriate to study fatigue on low-energy-irradiated samples where the radiation damage is produced in a narrow depth of few to 10s of micrometers. FEM techniques should be investigated to extend fatigue studies on low-energy irradiated samples.

## 8.4 Development/use of alternatives for mechanical fatigue test

Thermal fatigue due to pulsed beam energy deposition can be replicated via conventional mechanical fatigue studies using a tensile machine. Such machines usually run at low frequency, requiring very long runs to obtain the enough experimental data to estimate the lifetime of material (S-N curve). Developing an alternative fatigue test machine that operates at a much higher frequency will drastically reduce the experimental data acquisition time during the R&D cycle. We will investigate new methods that were developed using thin-film bending or ultrasonic tests.

# 9 Post Irradiation Examination

Most of the Post Irradiation Examination (PIE) is defined and well developed. Nevertheless, a few improvements are needed to better support the specific material studies needed for HPT, as presented in this Section.

## 9.1 Timeline

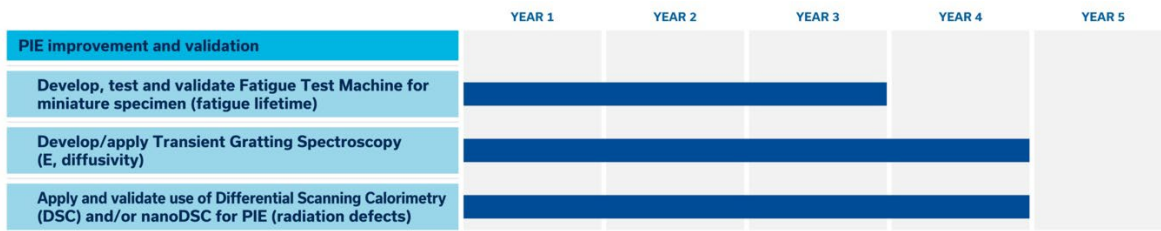

Figure 9 - Timeline for Post Irradiation Examination development to support HPT material R&D program.

We rely on extensive PIE capabilities at numerous different facilities to support the HPT R&D program. A list of these capabilities can be found in [47].

## 9.2 Fatigue test Machine for miniature specimen

The ability to perform testing and PIE on miniature specimens helps reduce experimental complexity and cost. Reducing specimen size also decreases activated volume and can facilitate the use of advanced characterization techniques located in radiological facilities outside of a hot cell. Conventional fatigue machines use standards with sizes up to a few inches. Development and validation of a fatigue test machine for miniature specimens will help to reduce the cost of the R&D cycle.

## 9.3 Develop and apply Transient Grating Spectroscopy (TGS) for irradiated material.

This advanced tool was developed and used at MIT as an ultrarapid, nondestructive materials evaluation technique [48]. Elastic, thermal, and acoustic properties allow users to probe material structures across multiple length scales. By generating surface acoustic waves and temperature gratings on a material, we can extract its elasticity, thermal diffusivity, and energy dissipation on a sub-microsecond scale. We would like to apply this technique and validate it for high-energy-proton and low-energy-ion irradiated materials for comparison.

## 9.4 Apply and validate the use of Differential Scanning Calorimetry and nanoDSC for PIE

Differential Scanning Calorimetry (DSC) is a thermomechanical technique to probe temperature and heat flow associated with thermal transitions in a material. NanoDSC allows testing of very small quantities of material and is typically used in conventional materials science. We will investigate the use of both tools to probe radiation defects in materials. If validated, nanoDSC will allow use of very small amounts of activated material to analyze radiation defects.

# 10 Facilities to support the R&D program.

None of these tests and studies can be done without specific facilities, including irradiation stations and PIE facilities. A non-exhaustive list of existing facilities that provide beams available for material studies in the context of HEP and High-Power-Targetry development can be found in [47].

## 10.1 Timeline

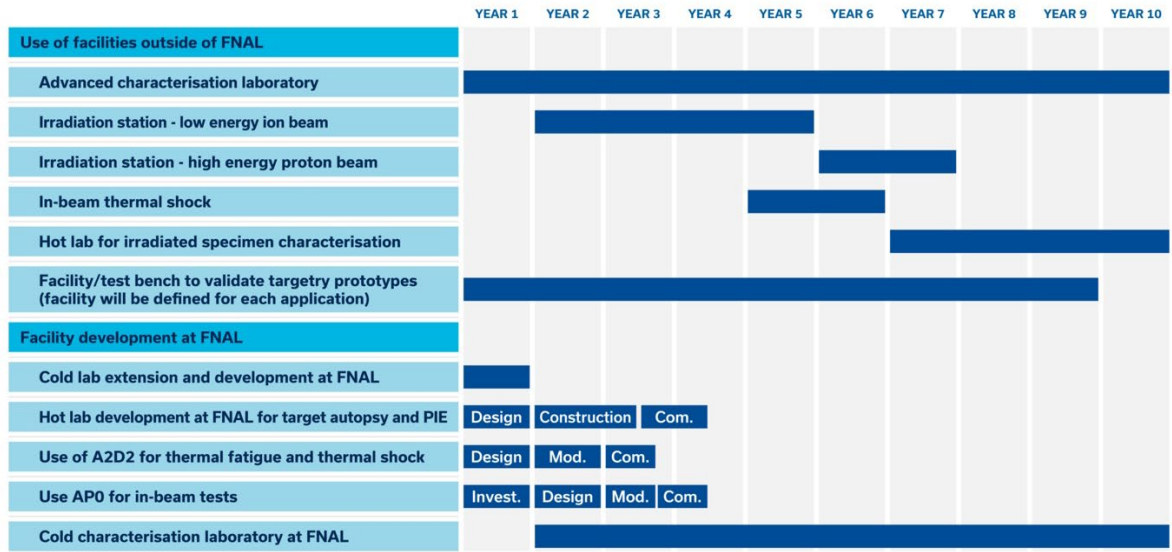

Figure 10 – List of facilities to be used in carrying out the HPT R&D program.

Each facility may have limited user access, and may partially meet the need for specific testing related to accelerator targetry development. Some PIE capabilities are limited or do not exist near some irradiation stations, leading to the need of transporting activated materials to other capable facilities, increasing both cost and duration of the R&D cycle.

To reduce R&D cost and shorten the duration of the R&D cycle, use of existing facilities at Fermilab will be investigated to assess their use for HPT R&D program.

## 10.2 A2D2 electron gun

The Accelerator Application Development and Demonstration (A2D2) research tool is a test platform using a 9-MeV electron gun at a maximum power of 1.2 kW. The need to add focusing at the gun exit is under investigation. If beam size is reduced to a sigma of ~1mm, this equipment could be used as the thermal shock test station and even perhaps for in-beam thermal fatigue testing of materials.

## 10.3 AP0

The AP0 target station currently hosts the Muon g-2 experiment. In the midterm, the possibility to use this facility as a general target test facility after Muon g-2 is complete needs to be investigated. This station offers the possibility of in-beam thermal and structural testing for targets as well as creation of irradiation damage in materials at a damage dose high enough to identify any early failures caused by induced changes in the materials.

## 10.4 Develop and extend the FNAL HPT lab.

The capabilities of FNAL's HPT testing laboratory are being extended rapidly. More space is needed to fully evaluate mechanical and physical properties of non-activated materials, and to develop and manufacture various novel materials.

To be able to perform rapid autopsies of targets that fail during operation and to avoid transport of activated materials to other capable facilities, it is critical to have PIE capabilities on-site at Fermilab, to limit cost and reduce the duration of the R&D cycle. Hot cells with limited PIE capability should be available to analyze/inspect irradiated materials (from target operation or irradiated material from irradiation campaigns) or irradiated beam intercepting devices. These hot cells should be flexible enough to accept a modest amount of characterization equipment. More advanced PIE techniques, at higher cost, will be used at other facilities.

# 11 Safety

During the R&D cycle, some studies will focus on materials that have been previously irradiated. These materials will be studied and tested to determine the effects of accelerator-based irradiation under different conditions on the physical and mechanical properties of target materials. The activated materials will come from two sources:

- The primary source will be samples extracted from current and future targetry components from HEP accelerators. Small material samples (on the order of 1 – 10 cm) can be extracted from horn, production target, beam windows, collimators, or any other HEP beam-intercepting devices.
- The second source of activated samples will be specimens fabricated for dedicated irradiation in support of the HPT R&D program. These specimens will be small (approximately 5 cm or smaller) and fabricated specifically for mechanical and thermal testing of activated materials. These samples may be irradiated in specific facilities as described in Reference [47]. Irradiation conditions on the R&D samples irradiated at outside proton beam facilities will be similar to those expected during HEP operation, and the resulting isotopes produced will be the same as those currently produced at Fermilab, i.e. the fissile or alpha-emitting products from the second source of activated samples will not be produced.

Any activities with activated materials at chosen facility will follow ALARA procedures and any other regulations in place at the facility.

In previous multi-material irradiation campaigns at the Brookhaven Linac Isotope Producer (BLIP) facility, the maximum dose rate from the most activated capsule was calculated to be approximately 40 R/hr at 1 ft (37.1 mSv/hr at 1 m). This capsule contained a total of 56 specimens and was irradiated in two phases for a total of 55 days with $4.6 \times 10^{21}$ accumulated protons. In the future, the R&D plan doesn't anticipate use of the BLIP facility for long irradiation times. If other irradiation facilities will be used in the future, we don't anticipate higher doses than 40R/hr at 1ft, or changes to the isotope inventory in the irradiated material.

Limiting the dose rate to 40 R/hr at 1 ft and following the irradiation with a 3-month cool-down period will allow use of more-economical, non-exclusive shipment in a Type A container. Following these steps will allow us to meet the regulatory requirements of both the U.S. Nuclear Regulatory Commission (NRC) and Department of Transportation (DOT). If longer irradiation times are deemed necessary in the future, or irradiation will be performed at another facility, other irradiation parameters (such as thickness of the sample, size of the beam, and cool-down time) and shielding can be adjusted to maintain the limit of 40R/hr at 1 ft.

## 12 Synergy with other communities

The High-Power Targetry community has been collaborating and working on several R&D projects to address the target challenges through the High-Power Targetry Workshop. The major high-energy laboratories in North America, Asia, and Europe are represented in this HPT workshop that began in 2003, and through the RaDIATE collaboration [21], led by Fermilab, since 2013.

The RaDIATE collaboration was formed to address challenges due to radiation damage to materials by bringing together experts from the fields of nuclear materials (fission and fusion power) and accelerator target facilities. The collaboration has grown to 20 participating institutions globally. Current activities include post-irradiation examination of materials taken from existing beamlines, as well as new irradiations of candidate target materials at low-energy and high-energy beam facilities. In addition, the program includes thermal shock experiments utilizing high-intensity proton beam pulses available at the HiRadMat facility at CERN [49].

Except the EMIR&A [50] and RaDIATE networks in France and Europe, no such network with irradiation facilities and associated PIE exists in the US or is available to support the HEP HPT R&D program. These networks, based on a single portal, facilitate experiments and create a strong synergy between the users and the facilities.

A similar network within the Nuclear Science User Facilities (NSUF) mission [51] exists in the US to support nuclear science and technology by providing nuclear energy researchers with access to world-class capabilities at no cost to the researchers. Unfortunately, HEP researchers don't have access to this network.

These national or international collaborations and networks will be essential in the next 10 years to expand our knowledge on high power targets.

# 13 Appendix A: Charge to the 2023 HPT Roadmap workshop

# 14 Appendix B: List of Attendees of the 2023 HPT Roadmap workshop

| Name | Institution |
| --- | --- |
| Kavin AMMIGAN | Fermilab |
| Gaurav ARORA | Fermilab |
| Charlotte BARBIER | Oak Ridge National Laboratory |
| Sujit BIDHAR | Fermilab |
| Abraham BURLEIGH | Fermilab |
| Marco CALVIANI | European Laboratory for Particle Physics (CERN) |
| Andrew CASELLA | Pacific Northwest National Laboratory |
| Chris DENSHAM | Rutherford Appleton Laboratory |
| Patrick HURH | Fermilab |
| Dohyun KIM | Brookhaven National Laboratory |
| Dong LIU | University of Bristol |
| Derun LI | DOE SC, OHEP |
| Kevin LYNCH | Fermilab |
| Shunsuke MAKIMURA | High Energy Accelerator Research Organization |
| Frederique PELLEMOINE | Fermilab |
| David SENOR | Pacific Northwest National Laboratory |
| Vladimir SHILTSEV | Fermilab |
| Diktys STRATAKIS | Fermilab |
| Jeff TERRY | Illinois Institute of Technology |
| Katsuya YONEHARA | Fermilab |

## 15  Appendix C: Agenda of the 2023 HPT Roadmap workshop

### Tuesday, April 11, 2023

| | Time | Title | Speakers |
|---|---|---|---|
| | 8:00-8:30 AM | Welcome | LI Derun<br>PELLEMOINE Frederique |
| | 8:30-9:30 AM | HPT Roadmap - HEP context | LYNCH Kevin |
| | 9:30-10:30 AM | Novel material and Concepts for Next generation High Power Targetry | DENSHAM, Chris<br>AMMIGAN, Kavin<br>CALVIANI, Marco<br>HURH, Patrick |
| | 10:30-11:30 AM | Modeling Needs | BARBIER, Charlotte |
| | 11:30 AM-1:00 PM | LUNCH | |
| | 1:00-2:00 PM | Radiation hardened beam instrumentations for Multi-MW beam facilities | YONEHARA, Katsuya |
| | 2:00-3:00 PM | Irradiation station and alternative beam + PIE needs | DENSHAM, Chris<br>SENOR, David<br>STRATAKIS, Diktys<br>KIM, Dohyun<br>TERRY, Jeff<br>CALVIANI, Marco<br>HURH, Patrick<br>MAKIMURA, Shunsuke |
| | 3:00-4:00 PM | Any missing topics | |

### Wednesday, April 12, 2023

| | Time | Title | Speakers |
|---|---|---|---|
| | 8:30-9:30 AM | Review points from Day 1 | PELLEMOINE Frederique |
| | 9:30 AM-12:00 PM | report review and update | All participants |
| | 12:00-1:20 PM | LUNCH | |
| | 1:30-3:30 PM | report review and update | All participants |

# 16 Appendix D: Glossary

This document contains acronyms and other terms that might not be easily deciphered even by experts in other areas of physics. Many of these terms are defined below. Some acronyms are defined using other acronyms. In those cases, please find the definitions of those acronyms at the appropriate place in this list.

**ABP**: Accelerator and Beam Physics thrust in the **GARD** program of the **DOE**.

**ACE**: Fermilab Accelerator Complex Evolution

**AMF**: The Advanced Muon Facility is a proposal for a next-generation muon facility at **Fermilab**.

**ASIC**: Application-Specific Integrated Circuit.

**BIF monitor**: Beam Induced Fluorescence monitor.

**BLM**: Beam Loss Monitor.

**CERN**: Conseil Européen pour la Recherche Nucléaire, the major European high-energy physics laboratory, located in Geneva, Switzerland.

**CNT**: Carbon Nano Tube.

**DOE**: U.S. Department of Energy.

**DPF**: **APS** Division of Particles and Fields.

**DSC**: Differential Scanning Calorimetry.

**DUNE**: Deep Underground Neutrino Experiment, a next-generation long-baseline neutrino oscillation experiment, based at Fermilab and **SURF**.

**EMT**: Electron Multiplier Tubes.

**EPPSU**: European Particle Physics Strategy Update. The European Strategy for Particle Physics is the cornerstone of Europe's decision-making process for the long-term future of the field.

**FCC**: Future Circular Collider, a large circular collider project proposed for **CERN**.

**FCC-ee**: A large circular e+e− collider proposed as a phase of the **FCC** project.

**FCC-hh**: A large circular proton-proton collider proposed as a phase of the **FCC** project.

**Fermilab**, also **FNAL**: Fermi National Accelerator Laboratory, in Batavia, Illinois, USA.

**GARD**: General Accelerator Research and Development program of the **DOE**.

**HEA**: High Entropy Alloy

**HEP**: High-Energy Physics, the generic term for particle physics research.

**HEPAP**: High Energy Physics Advisory Panel.

**HL-LHC**: High-Luminosity Large Hadron Collider, project aims to crank up the performance of the **LHC** by increasing the integrated luminosity by a factor of 10 beyond the **LHC's** design value.

**HPT**: High Power Targetry; including any beam intercepting devices.

**HRS**: Heat and Radiation Shield.

**HIP**: Hot Isostatic Pressing.

**JAEA**: Japan Atomic Energy Agency, Ibaraki, Japan,

**J-PARC**: Japan Proton Accelerator Research Complex, join project between **KEK** and **JAEA**, located in Ibaraki, Japan.

**KEK**: Ko-Enerugi Kenkyusho, the major high-energy physics laboratory in Japan, located in Tsukuba.

**LBNF**: Long Baseline Neutrino Facility, two facilities—Near Site at **FNAL** and Far Site at SURF—which together furnish the beam, underground facilities, and infrastructure required to support **DUNE**.

**LDV**: Laser Doppler Vibrometers.

**LHC**: Large Hadron Collider, a large proton–proton collider at **CERN**, with design CM energy 14 TeV.

**MW**: Mega Watt

**NITE SiC/SiC** composite: Nano-powder Infiltration and Transient Eutectoid Silicon Carbide composite.

**NSF**: US National Science Foundation.

**NSUF**: Nuclear Science User Facilities is the U.S. Department of Energy Office of Nuclear Energy's only designated nuclear energy user facility.

**OHEP**: Office of High-Energy Physics of the US **DOE**.

**ORNL**: Oak Ridge National Laboratory, Tennessee, USA.

**OTR monitor**: Optical Transition Radiation monitor.

**PAR**: **PIP-II** era Accumulator Ring has been proposed to enable the PIP-II linac to better perform both roles; to improve Booster performance and to provide a platform for new **HEP** experiments.

**PIE**: Post Irradiation Examination.

**PIP-II**: Proton Improvement Plan II is an essential enhancement to the **Fermilab** accelerator complex, powering the world's most intense high-energy neutrino beam.

**PSI**: Paul Scherer Institute is a multi-disciplinary research institute, located in Villigen and Würenlingen in Switzerland.

**P5**: Particle Physics Project Prioritization Panel, an advisory subcommittee of the **HEPAP**.

**SEM**: Secondary Electron emission Monitor.

**SEY**: Secondary Emission Yield.

**Snowmass'21**: the US **HEP** community strategic planning exercise, organized by the **DPF**.

**SNS**: Spallation Neutron Source, at **ORNL**.

**SURF**: Sanford Underground Research Facility: An underground laboratory in the former Homestake Mine in Lead, South Dakota, USA.

**TFGR**: Toughened, Fine Grained, Recrystallized tungsten alloy.

**TGS**: Transient Grating Spectroscopy.

**VTR**: Versatile Test Reactor will be a one-of-a-kind scientific user facility capable of performing large-scale, fast-spectrum neutron irradiation tests and experiments, located in Idaho, USA.

# 17 References.

[1] Accelerator and Beam Physics RoadMap, https://www.google.com/url?sa=t&rct=j&q=&esrc=s&source=web&cd=&cad=rja&uact=8&ved=2ahUKEwiZ8Nji1ZuBAxXdNzQIHVaBAWcQFnoECBsQAQ&url=https%3A%2F%2Fscience.osti.gov%2Fhep%2F-%2Fmedia%2Fhep%2Fpdf%2F2022%2FABP_Roadmap_2023_final.pdf&usg=AOvVaw1OyzXjkqGbt0yHhoByIlfq&opi=89978449

[2] LBNF – Long Baseline Neutrino Facility. DUNE at LBNF (fnal.gov)

[3] Falcone A., DUNE Collaboration, "Deep underground neutrino experiment: DUNE", Nuclear Instruments and Methods in Physics Research Section A, vol. 1041, 167217, 2022

[4] Valishev A., "Fermilab Proton Accelerator Complex Evolution (ACE) plan", Presentation at P5 Town Hall at Fermilab abd Argonne, March 21-24 2023. https://indico.fnal.gov/event/58272/

[5] Abada A., FCC Collaboration, "FCC-ee: The Lepton Collider: Future Circular Collider Conceptual Design Report Volume 2", Eur. Phys. J. Special Topics 228, 261-623 (2019). https://doi.org/10.1140/epjst/e2019-900045-4

[6] Abada A., FCC Collaboration, "FCC-ee: The Hadron Collider: Future Circular Collider Conceptual Design Report Volume 3", Eur. Phys. J. Special Topics 228, 755-1107 (2019). https://doi.org/10.1140/epjst/e2019-900087-0

[7] Byrum K. et al., "Mu2e-II: Muon to electron conversion with PIP-II", Snowmass 2022 white paper. https://doi.org/10.48550/arXiv.2203.07569

[8] Stratakis D., Mokhov N., Palmer M., Pastrone N., Raubenheimer T., Rogers C., Schulte D., Shiltsev V., Tang J., Yamamoto A., et al., "A Muon Collider Facility for Physics Discovery", 2022, https://arxiv.org/abs/2203.08033.

[9] Aoki M. et al., "A New Charged Lepton Flavor Violation Program at Fermilab", Snowmass 2022 white paper. https://doi.org/10.48550/arXiv.2203.08278

[10] W. Pellico et al., "FNAL PIP-II Accumulator Ring", Accelerator Physics, 2022, https://doi.org/10.48550/arXiv.2203.07339

[11] K. Ammigan et al., "Novel Materials and Concepts for Next-Generation Targetry Applications", Snowmass 2022 white paper. https://doi.org/10.48550/arXiv.2203.08357

[12] Kirk H.G. et al., "Target Studies with BNL E951 at the AGS", paper TPAH137 contributed to PAC2001 (June 18,2001)

[13] Ethymiopoulos I.; Fabich A.; Palm M.; Lettry J.; Haug F.; Pernegger H. et al., "The MERIT (nTOF-11) High Intensity Liquid Mercury Target Experiment at the CERN PS", article, June 23, 2008; United States. (https://digital.library.unt.edu/ark:/67531/metadc893165/ : accessed September 27, 2023), University of North Texas Libraries, UNT Digital Library, https://digital.library.unt.edu; crediting UNT Libraries Government Documents Department.

[14] Neuffer D. et al., "Pion-Production Target for Mu2e-II: Simulation Design and Prototype", Presented at the 23rd International Workshop on Neutrinos from Accelerators, Salt Lake City, UT, USA, 30–31 July 2022. Phys. Sci. Forum 2023, 8(1), 59; https://doi.org/10.3390/psf2023008059

[15] Thomason, J.W.G., "The ISIS Spallation Neutron and Muon Source—The First Thirty-Three Years", Nuclear Instruments and Methods in Physics Research Section A, vol. 917, pp. 61-67, 2018.

[16] Ahdida C. et al., "The Experimental Facility for the Search for Hidden Particles at the CERN SPS," Journal of Instrumentation, vol. 14, no. 3, pp. P03025-P03025, 2019.

[17] Nelson A.T. et al., Fabrication of a Tantalum-Clad Tungsten Target for LANSCE, Journal of Nuclear Materials, vol. 431, no. 1, pp. 172-184, 2012.

[18] Ahdida C. et al., SPS Beam Dump Facility - Comprehensive Design Study, arXiv:1912.06356, CERN-PBC-REPORT-2019-005, CERN-2020-002: https://cds.cern.ch/record/2703984?ln=en.

[19] Lopez Sola E., et al., "Design of a High Power Production Target for the Beam Dump Facility at CERN," Phys. Rev. Accel. Beams, vol. 22, p. 113001, 2019.

[20] Busom J. et al., "Application of Hot Isostatic Pressing (HIP) Technology to Diffusion Bond Refractory Metals for Proton Beam Targets and Absorbers at CERN," Mater. Des. Process. Commun., vol. 2, p. e101, 2019.

[21] A. Ahmad, C. Booth, D. Jenkins and T. Edgecock, "Generic Study on the Design and Operation of High-Power Targets," Physical Review Special Topics – Accelerator and Beams, vol. 17, 2014

[22] RaDIATE Collaboration (Radiaton Damage In Accelerator Target Environments), https://radiate.fnal.gov/

[23] B. Cantor et al., Microstructural Development in Equiatomic Multicomponent Alloys, Materials Science and Engineering: A, vol. 375-377, pp.213-218, 2004.

[24] Yeh J.W. et al., Nanostructured High-Entropy Alloys with Multiple Principal Elements: Novel Alloy Design Concepts and Outcomes, Advanced Engineering Materials, vol. 6, pp. 299-303, 2004.

[25] Ullah M. et al., Damage Accumulation in Ion-Irradiated Ni-based Concentrated Solid-Solution Alloys, Acta Materialia, vol. 109, pp. 17-22, 2016.

[26] Yang L. et al., High He-ion Irradiation Resistance of CrMnFeCoNi High-Entropy Alloy Revealed by Comparison Study with Ni and 304SS, Journal of Materials Science and Technology, vol. 35, pp. 300-305, 2019.

[27] Tong Y. et al., Evolution of Local Lattice Distortion under Irradiation in Medium-and High-Entropy Alloys, Materialia, vol. 2, pp. 78-81, 2018.

[28] El-Atwani, O et al., Outstanding Radiation Resistance of Tungsten-Based High-Entropy Alloys, Science Advances, vol. 5, no. 3, 2019.

[29] Lu C. et al, Enhancing Radiation Tolerance by Controlling Defect Mobility and Migration Pathways in Multicomponent Single-Phase Alloys, Nature Communications, vol. 7, 2016.

[30] Jin K. et al., Effects of Compositional Complexity on the Ion-Irradiation Induced Swelling and Hardening in Ni-containing Equiatomic alloys, Scriptia Materialia, vol. 119, pp. 65-70, 2016.

[31] Bidhar S. et al., Production and qualification of an electrospun ceramic nanofiber material as a candidate future high power target, Physical Review Accelerators and Beams, vol. 24, 123001, 2021

[32] Makimura S. et al., Tungsten Alloy Development as Advanced Target Material for High-Power Proton Accelerator, JPS Conf. Proc., vol. 18, 031002, 2020.

[33] Linsmeier Ch. et al., Development of Advanced High Heat Flux and Plasma-Facing Materials, Nucl. Fusion, vol. 57, 092007, 2017.

[34] Kurushita H. et al., Development of Nanostructured Tungsten Based Materials Resistant to Recrystallization and/or Radiation Induced Embrittlement, Mater. Trans., vol 54, no. 4, pp.456-465, 2013.

[35] Makimura S. et al., Development of Toughened, Fine Grained, Recrystallized W-1.1%TiC, Materials Science Forum, Spallation Materials Technology, vol. 1024, pp. 103-109, 2021.

[36] Kohyama A., et al., IP Conf. Ser. Mater. Sci. Eng., vol. 18, 202002, 2011.

[37] Makimura S., et al., Feasibility Study for NITE SiC/SiC as the Target Material for Pions/Muons Production at High-Power Proton Accelerator Facilities, JPS Conf. Proc., vol.28, 031005, 2020.

[38] Park J.S., et al., Ceramics International, vol. 44, 2018.

[39] Maestre J. et al, Journal of Instrumentation, vol 17, P01019, 2022.

[40] Nuiry F.X. et al., 3D Carbon/Carbon Composite for Beam Intercepting Devices at CERN, Material Design & Processing Communications, 1, e33, 2019.

[41] Barbier C. et al., Modeling Needs for High Power Target, Snowmass 2022 white paper. https://doi.org/10.48550/arXiv.2203.04714

[42] Yonehara K., Radiation hardened beam instrumentations for multi-Mega-Watt beam facilities, Snowmass 2022 white paper. https://doi.org/10.48550/arXiv.2203.06024

[43] Yonehara K. et al., Radiation robust rf gas beam detector R&D for intensity frontier experiments, FERMILAB-CONF-19-796-AD

[44] Ashida Y. et al., A new electron-multiplier-tube-based beam monitor for muon monitoring at the T2K experiment, Prog. Theor. Exp. Phys. 2018, 103H01

[45] Girard S. et al., Recent advances in radiation-hardened fiber-based technologies for space applications, J. Opt. 20 (2018) 093001.

[46] Barbeau P., Merkel P., and Zhang J., Report of the Instrumentation Frontier Working Group for Snowmass 2022 white paper, https://arxiv.org/pdf/2209.14111.pdf

[47] Baev, V.G., Vdovin, V.A., Vikharev, A.A. et al. Applied research using a 30 GHz free-electron maser: Experimental study of interacton of high-power pulsed radiation with metals. Radiophys Quantum El 54, 648–654 (2012). https://doi.org/10.1007/s11141-012-9310-3

[48] Pellemoine F. et al., Irradiation Facilities and Irradiation Methods for High Power Target, SnowMass 2022 white paper, https://doi.org/10.48550/arXiv.2203.08239

[49] Hofmann F., Short M.P., Dennett C. A., "Transient grating spectroscopy: An ultrarapid, nondestructive materials evaluation technique", https://arxiv.org/ftp/arxiv/papers/1908/1908.02051.pdf

[50] The HiRadMat Facility- High Radiation to Materials. https://hiradmat.web.cern.ch/

[51] Emir&a, French network of accelerators for irradiation and analysis of molecules and materials. https://emira.in2p3.fr/

[52] Rapid Turnaround Experiment at Nuclear Science User Facilities. https://nsuf.inl.gov/Page/rte